\documentclass[aps,superscriptaddress,reprint]{revtex4-1}

\usepackage{graphicx}
\usepackage{bm}
\usepackage{fancyref}
\usepackage{soul}
\usepackage[colorlinks,
	linkcolor=blue,
	urlcolor=blue,
	citecolor=blue]{hyperref}
\usepackage[T1]{fontenc}       
\usepackage[version=3]{mhchem}
\usepackage{xcolor}
\usepackage{titlesec}
\usepackage{pdfpages}
\usepackage{pgffor}
\titlespacing*{\section}{0pt}{1.1\baselineskip}{\baselineskip}

\usepackage{etoolbox} 

\makeatletter
\patchcmd{\@outputpage@head}{\@ifx{\LS@rot\@undefined}{}{\LS@rot}}{}{}{}
\makeatother

\begin{document}
	
	\title[]{Selective high frequency mechanical actuation driven by the \ce{VO2} electronic instability}
	
	\author{Nicola \surname{Manca}}
	\email{n.manca@tudelft.nl}
	\email{nicola.manca@spin.cnr.it}
	\affiliation{Kavli Institute of Nanoscience, Delft University of Technology, Lorentzweg 1, 2628 CJ Delft, The Netherlands}
	
	\author{Luca \surname{Pellegrino}}
	\affiliation{CNR-SPIN, Corso Perrone 24, 16152 Genova, Italy}
	
	\author{Teruo \surname{Kanki}}
	\affiliation{Institute of Scientific and Industrial Research, Osaka University, Ibaraki, Osaka 567-0047, Japan}
	
	\author{Warner J. \surname{Venstra}}
	\affiliation{Kavli Institute of Nanoscience, Delft University of Technology, Lorentzweg 1, 2628 CJ Delft, The Netherlands}
	
	\author{Giordano \surname{Mattoni}}
	\affiliation{Kavli Institute of Nanoscience, Delft University of Technology, Lorentzweg 1, 2628 CJ Delft, The Netherlands}
	
	\author{Yoshiyuki \surname{Higuchi}}
	\affiliation{Institute of Scientific and Industrial Research, Osaka University, Ibaraki, Osaka 567-0047, Japan}
	
	\author{Hidekazu \surname{Tanaka}}
	\affiliation{Institute of Scientific and Industrial Research, Osaka University, Ibaraki, Osaka 567-0047, Japan}
	
	\author{Andrea D. \surname{Caviglia}}
	\affiliation{Kavli Institute of Nanoscience, Delft University of Technology, Lorentzweg 1, 2628 CJ Delft, The Netherlands}
	
	\author{Daniele \surname{Marr\'e}}
	\affiliation{CNR-SPIN, Corso Perrone 24, 16152 Genova, Italy}
	\affiliation{Physics Department, University of Genova, Via Dodecaneso 33, 16146 Genova, Italy}
	
	\date{\today}
	
	\begin{abstract}
		\noindent
		\textbf{This is the pre-peer reviewed version of the following article: ``N. Manca et al. Adv. Mater. 29 (4), 1701618 (2017)'', which has been published in final form at: \url{http://doi.wiley.com/10.1002/adma.201701618}.
		This article may be used for non-commercial purposes in accordance with Wiley Terms
		and Conditions for Self-Archiving.}	
	\end{abstract}
		
	\maketitle

Ultra-thin free-standing structures such as membranes or microbridges can efficiently couple their mechanical degrees of freedom to electronic, optical and magnetic interactions in different excitation/response schemes~\cite{Bunch2007a,VanThourhout2010,Riedinger2016,Castellanos-Gomez2015}.
The use of materials with intrinsic functionalities and complex response to external stimuli constitutes the keystone towards next-generation miniaturized sensors and actuators~\cite{Masmanidis2007b,Park2011,Grover2011,Tao2014}.
To this purpose, transition metal oxides are a unique class of materials, where the balance between electronic correlations, magnetic ordering and lattice distortions gives rise to phase transitions and non-linear behaviours~\cite{Imada1998,Morosan2012}.
Among them, \ce{VO2} is considered a textbook example due to its metal-insulator transition associated with a crystal symmetry change when its temperature is increased above 65\,$^\circ$C~\cite{Morin1959a}.
The phase transition of \ce{VO2} is at the same time a puzzling mix of Mott physics, structural distortions~\cite{Eyert2002,Budai2014a,Brockman2014,Kumar2014} and a unique candidate for a variety of technological applications~\cite{Driscoll2009,Kats2014a,Pellegrino2012,Shukla2015,Xiao2015,Yoon2016d}.
One of the most fascinating characteristics of \ce{VO2} is the possibility of realizing a current/voltage periodic instability under constant electrical bias that determines electrical oscillations.
Spontaneous oscillations are a hallmark of non-linear systems~\cite{Jenkins2013}, and in \ce{VO2} they are triggered by the strong non-linear variation of its electrical properties across the phase transition.
This oscillating state has been investigated in \ce{VO2} bulk crystals~\cite{Taketa1975,Fisher1978} and more recently in single-crystal nano-beams and thin films~\cite{Gu2007,Lee2008a}.
Several studies showed how it is possible to control the frequency and the onset of this oscillating state by external parameters, such as device geometry, electrical bias, laser heating or by connecting electrical passive elements~\cite{Kim2010,Seo2011,Kim2012,Wang}.
So far all these studies focused on the analysis of the electrical characteristics of this oscillating state, considering it a potential platform for neural-mimicking computing architectures~\cite{Datta2014a,Shukla2014,Beaumont2014,Pergament2015a}.
However, since the phase transition of \ce{VO2} involves both its electronic and lattice properties, electrical oscillations shall also determine a strong periodic mechanical forces on the device structure, whose implications have not been studied so far.

Here, we demonstrate how the coupled resistive and structural transition of \ce{VO2} can be employed to generate mechanical excitations in the MHz range using only a DC voltage source. This is a local self-actuation mechanism capable of driving the motion of a micro-mechanical resonator, performing a direct transduction from a voltage bias to high-frequency mechanical excitation which relies on the intrinsic properties of \ce{VO2} only. In contrast to the typical approaches of mechanical actuation~\cite{Feng2008,Unterreithmeier2009,Baek2011}, our scheme is intrinsic to the device and does not require dedicated driving electronics, reducing device complexity, size and power consumption.

\begin{figure*}
	\includegraphics[width=\linewidth]{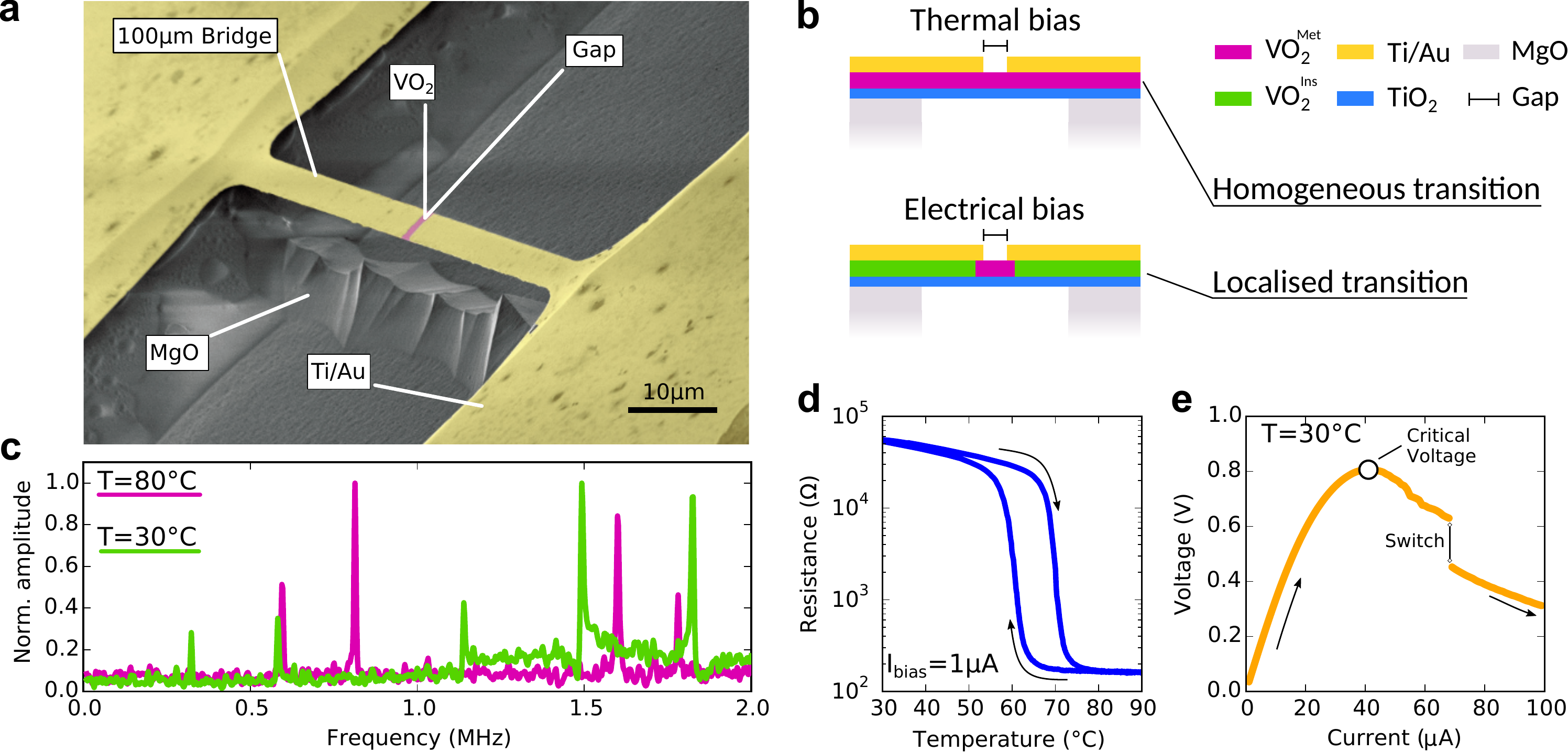}
	\caption{\label{fig:figure1}
		\textbf{Device structure and characterization.}
		(a) False-coloured SEM picture of a microbridge showing the \ce{MgO} substrate, the \ce{Ti/Au} electrodes and the 2\,$\mu$m gap of exposed \ce{VO2}.
		(b) Schematic side-view of the microbridge heterostructure illustrating the homogeneous phase transition when the temperature is varied using an external heater (top) and the localized phase transition due to the confined Joule heating (bottom).
		(c) Mechanical response of the microbridge at T = 30\,$^\circ$C and T = 80\,$^\circ$C taken at $\mathrm{10^{-4}}$\,mbar background pressure as measured with an optical detector. A distinct upward shift of the resonance frequencies is detected in the high-temperature phase.
		(d) Resistance vs temperature characteristics of the microbridge taken at low current bias. (e) Voltage vs current characteristics of the microbridge. The ``Critical Voltage'' indicates the point of electro-thermal runaway if under voltage bias, while ``Switch'' indicates the localized phase transition to the metallic state of the \ce{VO2} within the gap.
	}
\end{figure*}

Our device is a $\mathrm{100\,\mu m \times 5\,\mu m}$ free-standing microbridge made of a 113\,nm-thick \ce{TiO2 / VO2} crystalline bilayer. Metallic \ce{Ti / Au} electrodes (50\,nm-thick) partially cover the structure and provide good electrical contact (Fig.\,\ref{fig:figure1}a).
The \ce{TiO2 / VO2} heterostructure is deposited by Pulsed Laser Deposition as described in the Methods. The resulting lattice is c-oriented with domains having orthogonal in-plane orientations~\cite{Pellegrino2012,Okimura2005,Muraoka2002a}.
Details of the fabrication process can be found in the Methods section and in Supplementary section I, while further aspects of the device are discussed in the Supplementary section II.
The phase transition (PT) of \ce{VO2} is triggered above 64\,$^\circ$C, where its electrical resistivity drops by more than two orders of magnitude (insulator-metal transition) and the lattice symmetry changes from monoclinic to rutile (structural transition).
Both effects are observed in the microbridge when it is homogeneously heated by an external thermal bias (Fig.\,\ref{fig:figure1}b, top): the full bridge undergoes the PT, the lattice transformation causes a frequency shift of the mechanical modes up to 50\,\% (Fig.\,\ref{fig:figure1}c and Supplementary section III) and the electrical resistance shows its characteristic hysteresis loops (Fig.1d).
This frequency shift is due to the different in-plane lattice constants of \ce{VO2} between the monoclinic and rutile phases~\cite{Eyert2002}, that results in an isotropic strain of 0.08\,\% in the planar directions (See Supplementary section IV).
As a consequence, the device experiences the stress-stiffening~\cite{Karabalin2012} and geometric-stiffening~\cite{Pini2016} effects, where the first is related to an increase of the tension within the structure and the second is a geometric deformation which both increase the structure rigidity.
This wide variation of the internal mechanical stress across the phase transition has been recently exploited to realize tunable MEMS, like programmable resonators and static actuators~\cite{Rua2010a,Wang2013,Manca2013,Liu2014a,Merced2015,Ma2016}.
We note that the mechanical spectra of the microresonator, reported in Fig.\,\ref{fig:figure1}c, differ from that of the simple double-clamped beam model.
This because, as discussed in Supplementary section II, the device profile has a slightly buckled shape due to the built-in compressive stress within the heterostructure.
In Supplementary section V we show finite elements simulations of the mechanical modes of our microbridge under buckling conditions and how strain can determine the measured frequency shifts.

In order to excite the resonator without altering its mechanical spectrum, we designed \ce{Ti / Au} electrodes on top of the free-standing structure that leave a small active \ce{VO2} region just within their gap, as indicated in Fig.\,\ref{fig:figure1}a.
Since the electrical and thermal resistivity of gold are negligible compared to \ce{VO2}, these electrodes localize the voltage drop on the \ce{VO2} gap only and enhance its thermal contact with the clamped region.
An electrical bias applied to the microbridge thus results in heating by Joule effect which is localized in the \ce{VO2} gap, eventually triggering its PT.
Here, the metal electrodes play the crucial role of limiting the spread of the \ce{VO2} metallic phase (Fig.\,\ref{fig:figure1}b, bottom), making the mechanical spectrum independent from the electrical current (See Supplementary section III).
The voltage-current relationship $V(I)$ of the microbridge is plotted in Fig.\,\ref{fig:figure1}e, showing how the characteristic features given by \ce{VO2} are retained. The non-monotonic behaviour with a negative differential resistance observed above 40\,$\mathrm{\mu}$A and the sharp jump at 60\,$\mathrm{\mu}$A are the key elements for the realization of electrical oscillations, as will be discussed below.

\begin{figure}
	\includegraphics[width=\linewidth]{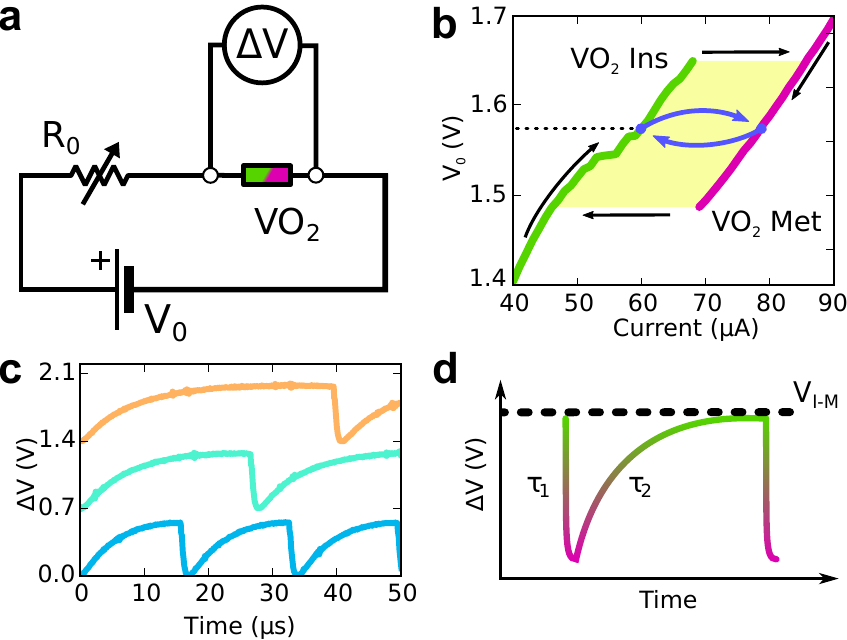}
	\caption{\label{fig:figure2}
		\textbf{Electrical oscillations under DC bias.}
		(a) Electrical circuit used to characterize the electrical oscillation. $V_{\mathrm{0}}$ is the applied DC voltage, $R_{\mathrm{0}}$ the variable load resistor.
		(b) Voltage-current characteristic of the circuit in (a), with $R_{\mathrm{0}}=15\,\mathrm{k\Omega}$. The Joule effect triggers the PT of \ce{VO2}, resulting in a hysteresis loop (black arrows). The yellow region indicates the voltage range where multiple current values are allowed. Under voltage bias (dashed line), specific combinations of $V_{\mathrm{0}}$ and $R_{\mathrm{0}}$ make the system oscillating, with a continuous switching between the two states (blue arrows).
		(c) Tuning of the oscillation period by varying $R_{\mathrm{0}}$ for $R_{\mathrm{0}} = 1.5\,\mathrm{V}$. From top to bottom: 16.25\,k$\mathrm{\Omega}$ (orange), 13.75\,k$\mathrm{\Omega}$ (green), 10\,k$\mathrm{\Omega}$ (blue). Traces are shifted by 0.7\,V for clarity.
		(d) The time constants of the two half periods, $\tau_1$ and $\tau_2$, are given by the different relaxation times of the metallic and insulating states, respectively. The switching from insulating to metallic state is triggered when the voltage drop across the \ce{VO2} gap ($V_{\mathrm{I-M}} = 0.6\,\mathrm{V}$) reaches a critical value, which also depends on the device geometry.
	}
\end{figure}

A consequence of the non-linear $V(I)$ of \ce{VO2} is the onset of electrical oscillations under DC bias. This state has been obtained using the circuit of Fig.\,\ref{fig:figure2}a, where a DC voltage source ($V_{\mathrm{0}}$) is applied to the \ce{VO2} microbridge and a load resistor ($R_{\mathrm{0}}$).
Their total voltage-current characteristic is reported in Fig.\,\ref{fig:figure2}b and illustrates the origin of the electrical oscillations in the circuit.
By sweeping the current magnitude, it is possible to track the hysteresis in the $V(I)$ relationship  due to the PT in the \ce{VO2}. At around 70\,$\mu$A the \ce{VO2} switches from insulating (green) to metallic (magenta) state, with lower electrical resistance.
When the circuit of Fig.\,\ref{fig:figure2}a is biased with a DC voltage, the hysteresis window determines a range of values where two current conditions are allowed (yellow region).
Under a constant voltage bias (dashed line), the voltage drop across the microbridge depends on the metallic or insulating state of \ce{VO2}.
In the insulating state the voltage is localized across the microbridge by choosing the values of $R_{\mathrm{0}}$ well below the electrical resistance of the \ce{VO2} element.
The resulting temperature increase can trigger the PT, lowering the microbridge electrical resistance.
In this case the voltage drop is mainly across the load resistor and the Joule heating on the microbridge is reduced, the temperature decreases and the initial insulating state is recovered starting a new cycle.
This continuous switching is a relaxation-oscillation condition~\cite{Jenkins2013}, and is triggered for specific combinations of $R_{\mathrm{0}}$ and $V_{\mathrm{0}}$~\cite{Maffezzoni2015}.
Fig.\,\ref{fig:figure2}c shows the voltage drop across the microbridge as a function of time while the electrical oscillations are triggered for $V_{\mathrm{0}} = 1.5\,\mathrm{V}$ and different $R_{\mathrm{0}}$ between 10\,k$\Omega$ and 20\,k$\Omega$.
These electrical oscillations have a double-exponential wave-shape (Fig.\,\ref{fig:figure2}d), where the time constants $\tau_1$ and $\tau_2$ of the semi-periods are determined by the thermal and electrical RC constants of the system.
Their amplitude is constant and is determined by the critical voltage for the insulator-to-metal transitions of the \ce{VO2} element ($V_{\mathrm{I-M}}$ of Fig.\,\ref{fig:figure2}d).
The electrical oscillation frequency ($f_{\mathrm{EO}}$), instead, can be controlled from 20\,kHz to 150\,kHz by adjusting $R_{\mathrm{0}}$ and $V_{\mathrm{0}}$.

\begin{figure}[b]
	\includegraphics[width=\linewidth]{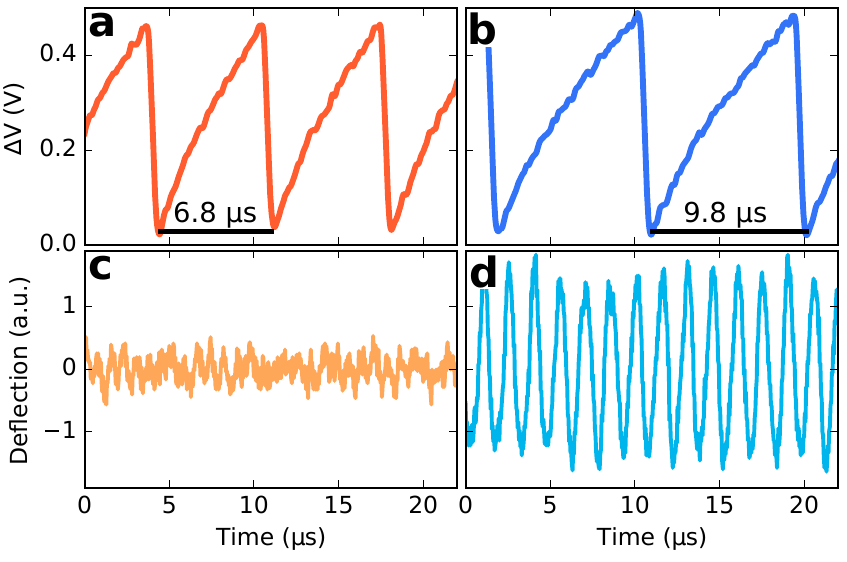}
	\caption{\label{fig:figure3}
		\textbf{Coupling electrical oscillation to mechanical motion.}
		(a) and (b) show time-domain electrical oscillations measured for two different periods of the electrical relaxation oscillation.
		(c) and (d) show the corresponding deflection signal measured at the same time by the photodiode.
		The trace in (c) has small amplitude and no clear harmonic component, while (d) shows that, for specific values of $f_{\mathrm{EO}}$, a strong periodic signal is detected.
		These traces were acquired by keeping a constant $V_{\mathrm{0}} = 2\,\mathrm{V}$ and $R_{\mathrm{0}}$ in the 10--20\,k$\mathrm{\Omega}$ range.
	}
\end{figure}

By tuning $f_{\mathrm{EO}}$ it is possible to excite the mechanical motion of the microbridge.
This is detected by measuring the deflection of the device with an optical lever setup, while the DC circuit is set to trigger the electrical oscillations (see Supplementary section VI).
Fig.\,\ref{fig:figure3} shows the time plots of the voltage drop measured across the microbridge (a, b)and its corresponding mechanical deflection (c, d) while the electrical oscillations are present.
For a generic combination of $R_{\mathrm{0}}$ and $V_{\mathrm{0}}$, the mechanical motion has small amplitude and no dominant harmonic component (a, c).
However, for specific values of these parameters, and consequently specific $f_{\mathrm{EO}}$, a strong harmonic mechanical response is observed at frequencies much higher than the electrical one (b, d).
This response corresponds to a mechanical resonance of the microbridge as measured in Fig.\,\ref{fig:figure1}c, meaning that, for specific $f_{\mathrm{EO}}$ values, the low-frequency electrical oscillations trigger resonant motion of the microbridge well above $f_{\mathrm{EO}}$.
As discussed in the next section, this is possible because the relaxation oscillation is non-sinusoidal and thus containing higher frequencies harmonic components.

The coupled electronic and structural transition of \ce{VO2} make possible three parallel mechanisms to achieve mechanical actuation: actuation by structural transition, actuation by thermal expansion and electrostatic coupling.
The first mechanism comes from the \ce{VO2} lattice change during the PT, where the unit cell periodically expands and contracts when switching between the insulating/monoclinic and metallic/rutile phases.
Thermal excitation comes from the Joule heating produced by the oscillating current, which results in a periodic thermal expansion of the structure.
The electrostatic actuation originates from the periodic oscillation of the microbridge voltage with respect to ground which produces a net capacitive force coupled to the surrounding dielectric environment.
All these three components are periodically modulated by the \ce{VO2} PT and can contribute synergistically to the mechanical actuation with a periodic excitation at $f_{\mathrm{EO}}$ and its higher harmonics.
An evaluation of their magnitude is presented in Supplementary section VII.

\begin{figure}
	\includegraphics[width=\linewidth]{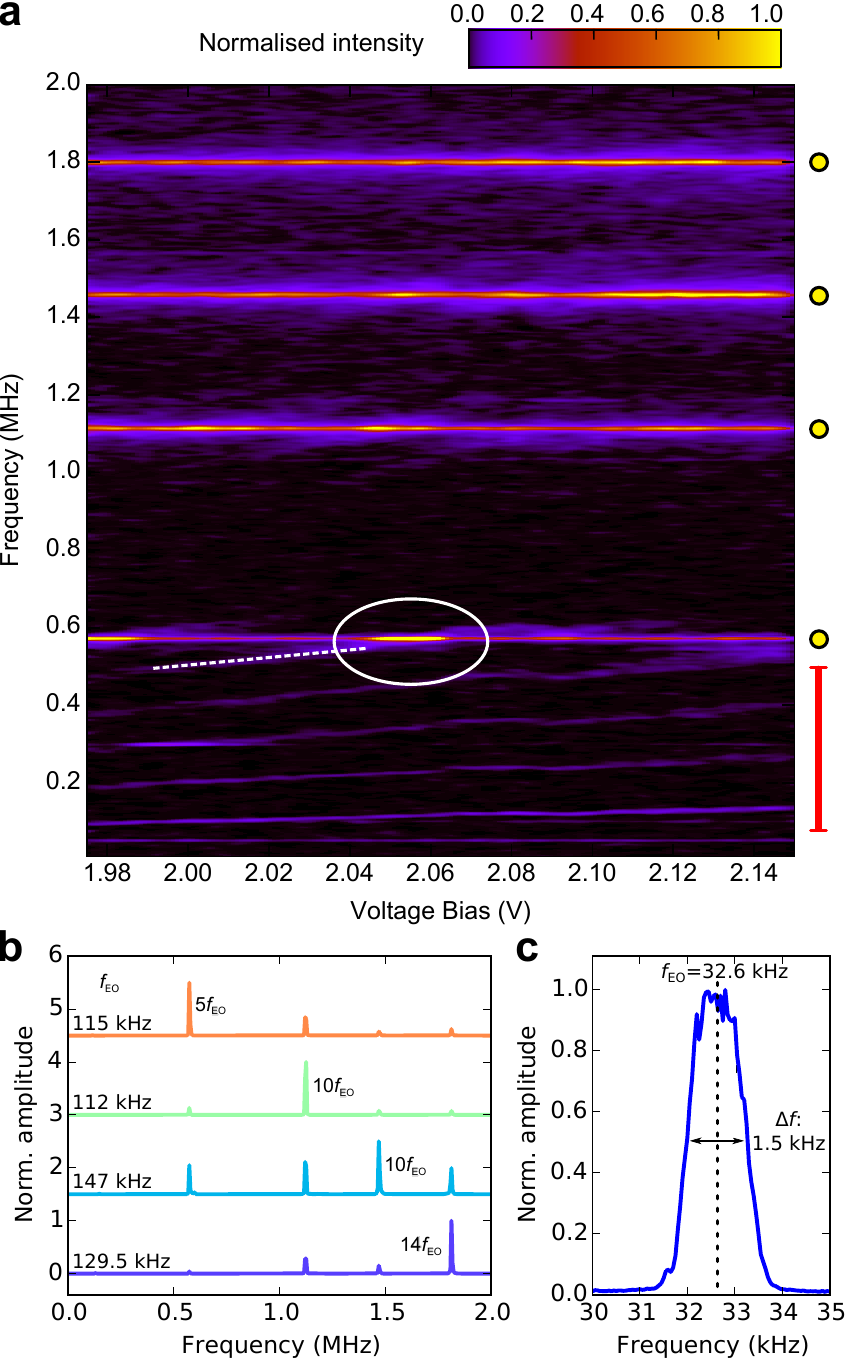}
	\caption{\label{fig:figure4}
		\textbf{Selective excitation of the mechanical resonance modes.}
		(a) Colormap showing the selective excitation of different mechanical modes of the microbridge by voltage bias: $\mathrm{R_0}$ is fixed at $20\,\mathrm{k\Omega}$ while $\mathrm{V_0}$ is varied. Resonant mechanical excitation is obtained when a harmonic component of the electrical oscillations matches a mechanical eigenfrequency of the microbridge (yellow dots).
		(b) Power spectral density of the photodiode signal taken at four different $f_{\mathrm{EO}}$ values. The amplitude of a specific mode is enhanced when its frequency (575\,kHz, 1120\,kHz, 1470\,kHz and 1800\,kHz) matches with of one of the higher harmonics of the electrical signal. The harmonic number is indicated in the figure.
		(c) Frequency jitter of the electrical oscillation, measured by averaging 100 spectra acquired over 40\,s.
	}
\end{figure}

We now demonstrate how the excitation of the different mechanical modes of the microbridge can be controlled by the electrical bias.
Fig.\,\ref{fig:figure4}a shows the power spectra of the deflection of the device measured by the photodiode as a function of $V_{\mathrm{0}}$, and with fixed $R_{\mathrm{0}}\mathrm{=20\,k}\Omega$. In this configuration, $V_{\mathrm{0}}$ controls the relaxation-oscillation frequency $f_{\mathrm{EO}}$, making our device a spontaneous voltage-controlled electrical oscillator with a gain of about 250\,kHz/V.
In Fig.\,\ref{fig:figure4}a, a comb of dim peaks is visible at low frequencies, marked with a red bar.
This is an off-resonance response of the device, corresponding to the first few harmonics of the electrical signal of Fig.\,\ref{fig:figure2}c.
By tuning $V_{\mathrm{0}}$ it is possible to match the frequency of one of these harmonic components with a flexural mode of the microbridge (yellow dots).
This condition is realized, for example, at the crossing point indicated by the circle in Fig.\,\ref{fig:figure4}a, where the increase of amplitude of the specific flexural mode is obtained through resonant excitation by the fifth harmonic of the electrical signal, also visible as forced mechanical oscillation in the unmatched condition (white dashed line).
The selective excitation of different mechanical modes is achieved by tuning $f_{\mathrm{EO}}$ to obtain the desired matching condition, as shown in the spectra of Fig.\,\ref{fig:figure4}b.
The labels indicate the matched harmonic component (n$f_{\mathrm{EO}}$) and the frequency of the selected mode, where, as an example, the 575\,kHz peak is excited with the fifth harmonic component of $f_{\mathrm{EO}}$ = 115\,kHz.

An important characteristic of an oscillator is its frequency stability.
In our device the selectivity of the different mechanical modes is determined by the stability of the electrical oscillation.
Fig.\,\ref{fig:figure4}c is an average of 100 acquisitions of the electrical oscillations spectra taken over 40\,s for $f_{\mathrm{EO}}$= 32.6\,kHz.
Its width indicates that the amplitude of the frequency jitter of the electrical oscillations, i.e., their stability over time, is about 4-5\,\%.
This means that even in an unmatched condition, fluctuations of $f_{\mathrm{EO}}$ can trigger the resonant actuation of modes close to one of its harmonic components.
This explains why in Fig.\,\ref{fig:figure4}a and b the mechanical modes of the microbridge are slightly visible also in the unmatched condition: as the integration time of the network analyser is much longer than these fluctuations, the instrument shows the averaged mechanical response.
The magnitude of the frequency jitter is related to the robustness of the two switching conditions of the \ce{VO2} within the gap and may be improved by optimizing device design, growth methods or metal contacts.

In this last paragraph we evaluate the energy efficiency of our device, how it scales with the size of the active region and compare the presented approach with other actuation mechanisms based on self-oscillation in micro and nanostructures.
The total amount of energy provided to the system is the total Joule power $P_{\mathrm{tot}}$ dissipated in the circuit of Fig.\,\ref{fig:figure2}a while the electro-thermal oscillations are activated.
It can be evaluated by considering that the $V(I)$ characteristic of the circuit in Fig.\,\ref{fig:figure2}b has as upper limit:
${P_{\mathrm{tot}} = V_{\mathrm{0}}\cdot I_{\mathrm{EO}} = 1.6\,\mathrm{V} \cdot 80\,\mathrm{\mu A} = 130\,\mathrm{\mu W}}$.
Since the load resistor is comparable to the gap resistance in the insulating phase~\cite{Maffezzoni2015}, as a first approximation this power is evenly split between the microbridge and $R_{\mathrm{0}}$.
The power required to undergo a periodic phase transition of a \ce{VO2} volume is the product between frequency and heat absorbed by the transition:
${P_{\mathrm{PT}} = f_{\mathrm{EO}}(\Delta T c^{p}_{\mathrm{VO_2}}+ \lambda_{\mathrm{VO2}})\rho_{\mathrm{VO2}}}$,
where ${\rho_{\mathrm{VO2}} = 4.7\,\mathrm{g/m^3}}$ is the density of \ce{VO2}, ${\lambda_{\mathrm{VO2}} = 51\,\mathrm{J/g}}$ is the latent heat of the transition and ${c^{p}_{\mathrm{VO_2}} = 0.75\,\mathrm{J/K\cdot g}}$ is the specific heat of \ce{VO2} \cite{Cao2009e,Berglund1969}.
For a frequency of 100\,kHz, a temperature variation of 10\,K and a volume corresponding with the \ce{VO2} gap, the resulting average power is about $P_{\mathrm{PT}} = 30\,\mathrm{nW}$.
This value is orders of magnitude below the actual power dissipated in the microbridge, meaning that the power consumption of our device is dominated by thermal dissipations.
Several reports in literature discuss the energy efficiency and performances of \ce{VO2}-based static and quasi-static actuators~\cite{Wang2013e,Guo2013,Cabrera2014}.
Here, it is relevant to compare the device power consumption with other self-actuation mechanisms proposed so far.
In suspended membranes of 2D materials, for example, it is possible to trigger self-oscillations by focusing a diverging beam of laser light~\cite{Barton2012}.
However, in order to achieve the required light flux gradients to trigger a resonant excitation, several mW of power (>2\,mW) are needed.
Furthermore, scaling is limited by diffraction, complex optics are necessary and the actuation is not mode-selective.
A different approach relies on a self-biasing feedback loop, typically coupled with an electrostatic actuator~\cite{Feng2008,Chen2013a,Chen2016d}.
In this case the power consumption is dominated by the active electronic components in the feedback circuits.
Here, a single amplifying elements, as reported in the datasheets of the components used in the experiments, requires at least 1\,W of electrical power.
In this context, the total power consumption of our \ce{VO2}-based mechanical actuation mechanism is dramatically lower.
Also, since it depends on the volume of \ce{VO2} undergoing the phase transition, it scales with the device size, thus foreseeing an enhanced efficiency for nanoscale devices.

In conclusion, we realized a simple and flexible actuation scheme for high-frequency mechanical resonators based on a phase-change material.
The intrinsic physical properties of \ce{VO2} allow the direct conversion of a small DC voltage into a mechanical excitation in the MHz range without the need of an external driving circuit.
This device can be viewed as a spontaneous voltage-controlled oscillator, which is able to selectively excite the different mechanical modes of a microstructure by controlling the bias voltage.
Our actuation mechanism can be implemented in a variety of micro-electro-mechanical systems requiring resonant actuation, and the device size can be scaled down to comprise just a single sub-micrometric \ce{VO2} domain, allowing individual actuation of nanometric structures.
Our approach is thus scalable both in size and number of devices, where multiple micro-/nano-resonators can be actuated in parallel by a single DC source.
This opens the possibility of realizing DC-powered arrays of micro-actuators, fast frequency-switching devices or sensors based on multi-frequency detection, with potential applications spanning from micro-robotics to micro-fluidic devices and environmental monitoring.
	
\section*{Methods}
\paragraph*{Device Fabrication.}
The \ce{TiO2(23nm) / VO2(90nm)} heterostructure is grown on top of a \ce{MgO(100)} single-crystal substrate by Pulsed Laser Deposition.
During the growth the substrate temperature was kept at 450\,$^\circ$C and the laser fluency was 18\,mJ/cm$^2$.
\ce{TiO2} is deposited in 0.1\,Pa of \ce{O2} with a laser repetition rate of 3\,Hz, while \ce{VO2} in 0.95\,Pa of \ce{O2} with a repetition rate of 2\,Hz.
The device is patterned by e-beam lithography using Poly(methyl-methacrylate) resist.
\ce{Ti(5nm)/Au(45nm)} electrodes are deposited by thermal evaporation, followed by lift-off in acetone.
The bridge geometry is defined by Ar ion milling with an energy of 500\,eV and an ion flux of about 0.2\,mA/cm$^2$. The structures are suspended by selective etching of the \ce{MgO} substrate in \ce{H3PO4} (8.5\,\% aqueous solution).

\paragraph*{Measurement of mechanical spectrum.}
Microbridge motion is detected with a focused laser by an optical lever technique. The laser wavelength is 658\,nm and the focused spot size about 2-3\,$\mu$m, well within than the bridge width.
The spot is focused at about one fourth of the bridge length, this guarantees a maximum geometric gain of the optical lever.
The laser power is kept low in order to minimize additional heating by the laser.
This is checked before each measurement by monitoring the electrical resistance of the microbridge with and without the laser on the structure.
The device is placed in a vacuum chamber with a Peltier element and a Pt100 thermometer.
Sample temperature is controlled with a PID feedback loop and kept fixed at 30\,$^\circ$C unless indicated otherwise.
All the mechanical spectra have been measured at $\mathrm{10^{-4}}$\,mbar.
The mechanical spectrum of Fig.\,\ref{fig:figure1}c is measured using a Vector Network Analyser through thermal excitation, by sending a small AC current of $I_{\mathrm{0}}(1+\sin(\omega t))$, $I_{\mathrm{0}} = 5\,\mathrm{\mu A}$.
This current value is well below that needed to drive the phase transition of \ce{VO2} in the microbridge. 

\paragraph*{Reported data.}
Measurements reported in Fig.\,\ref{fig:figure3} and Fig.\,\ref{fig:figure4} have been acquired from different microbridges fabricated on the same sample.
Measurements reported in Fig.\,\ref{fig:figure4}a and Fig.\,\ref{fig:figure4}b have been acquired in different runs.
In Fig.\,\ref{fig:figure4}a a dim peak is visible at about 50\,kHz whose frequency is voltage independent.
This signal is due to electronic interference in the photodiode pre-amplifier, and does not affect the measurements.
The mechanical power spectra of Fig.\,\ref{fig:figure4}a are measured using a spectrum analyser from the signal from the photodiode while the electrical oscillations are present.
The average of 20 traces is shown, where each trace contains 801 data points, taken across a frequency span of 10\,kHz--2\,MHz, at a measurement bandwidth of 1\,kHz.

\section*{Acknowledgements}
This work was supported by the Executive programme of cooperation between Italy and Japan by the Italian Ministry of Foreign Affairs, the Dutch Foundation for Fundamental Research on Matter (FOM), a Grant-in-Aid for Scientific Research A (No.26246013), a Grant-in-Aid for Scientific Research B (No.16H03871) from the Japan Society for the Promotion of Science (JSPS).

\section*{Author Contributions}
N.M and L.P. conceived the experiment. N.M. designed, fabricated and characterized the devices. T.K and Y.H. prepared the \ce{VO2/TiO2} heterostructure. W.J.V. assisted with setting up the measurement setup. G.M. provided support in the device fabrication and acquired the SEM images.  N.M. wrote the manuscript with the help of L.P. and W.J.V.. H.T., A.D.C. and D.M. supervised the project. All authors discussed the results and commented on the manuscript. N.M. thanks A. Filippetti and D.J. Groenendijk for the useful comments on the manuscript.

\bibliographystyle{apsrev4-1}
\bibliography{library}
\newpage\newpage



\section*{Supplementary material}

# Selective high frequency mechanical actuation driven by the VO$_2$ electronic instability

## Supplemental Material

N. Manca,[1, *] L. Pellegrino,[2] T. Kanki,[3] W. J. Venstra,[1] G. Mattoni,[1]
Y. Higuchi,[3] H. Tanaka,[3] A. D. Caviglia,[1] and D. Marré[4, 2]

[1]*Kavli Institute of Nanoscience, Delft University of Technology,*
*P.O. Box 5046, 2600 GA Delft, The Netherlands*
[2]*CNR-SPIN, Corso Perrone 24, Genova 16152, Italy*
[3]*Institute of Scientific and Industrial Research,*
*Osaka University, Ibaraki, Osaka 567-0047, Japan*
[4]*Physics Department, University of Genova,*
*Via Dodecaneso 33, Genova 16146, Italy*

(Dated: August 1, 2017)

* n.manca@tudelft.nl

This supplemental material contains the following:

Supplementary section I: Fabrication process of the microbridges.

Supplementary section II: SEM and optical images of a device and measurement of the Q-factor.

Supplementary section III: Temperature and current frequency shift of the mechanical modes.

Supplementary section IV: Lattice directions and strain induced by the $VO_2$ phase transition

Supplementary section V: Finite elements simulation of the mechanical eigenmodes of the microbridge under compressive strain.

Supplementary section VII: Measurements setup.

Supplementary section VII: Comparison of the contributions to the electro-thermal actuation.

# I. Fabrication process

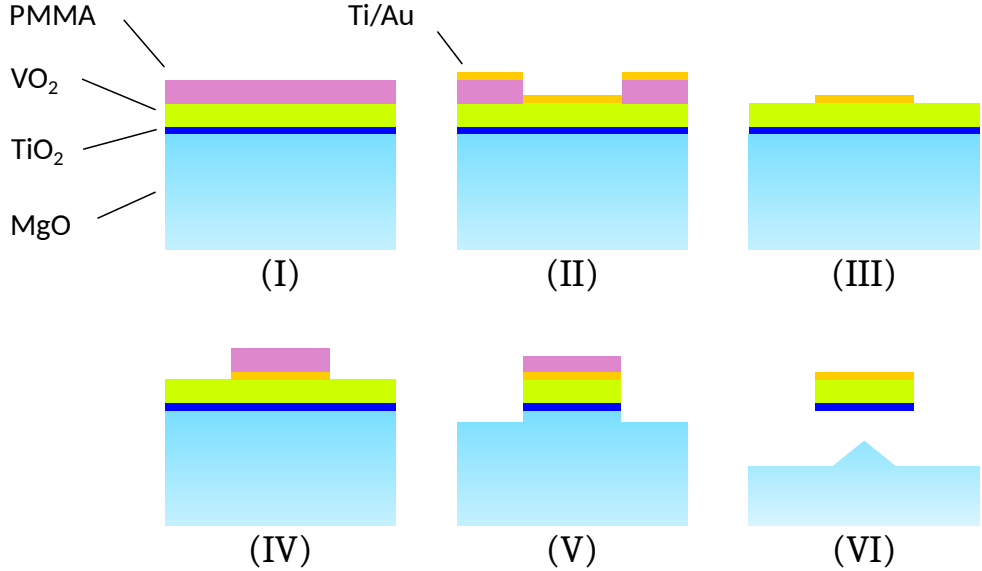

Figure S1. Fabrication process of the micromechanical resonator from VO$_2$/TiO$_2$ thin films.

Figure S1 shows a schematic of the main steps of the fabrication process of our device. We initially deposit by Pulsed Laser Deposition a heterostructure made of a 20 nm-thick TiO$_2$ thin film layer followed by a 90 nm-thick VO$_2$ thin film deposited on a 5 mm × 5 mm × 0.5 mm MgO(001) substrate. The microbridge is prepared from this heterostructure following the main steps listed below:

I We deposit by spin-coating a layer of Poly(methylmethacrylate) (PMMA) used as resist for E-beam lithography.

II Once the geometry of the metal contacts is realized, we deposit a 5 nm-thick sticking layer of Ti, followed by 45 nm of Au.

III The Au/Ti metal pad geometry is finally obtained by lift-off sonicating in acetone, kept at the temperature of 40 °C, for about 15 minutes.

IV The microbridge geometry is obtained by a second E-beam exposure, realizing a protective PMMA layer that covers the final device geometry.

V The exposed $VO_2/TiO_2$ layers are then etched by Ar+ ion milling. Ion energy is 500 eV and the current flux is calibrated to maintain an etching rate of about 20 nm/min. After the dry etching process the PMMA is removed with the same method described in point (III).

VI The final step is the release of the microbridge structure from the MgO substrate. This is done by soaking the sample in $H_3PO_4$ 8.5% water solution. The acid bath is kept at the constant temperature of 30 °C and continuously agitated with a magnetic stirrer at 100 RPM. After 3-4 hours a gap between the microbridge and the substrate is created. We finally dry our sample in a $CO_2$ Critical Point Dryer using ultra-pure isopropyl alcohol as buffer liquid.

## II. Final device

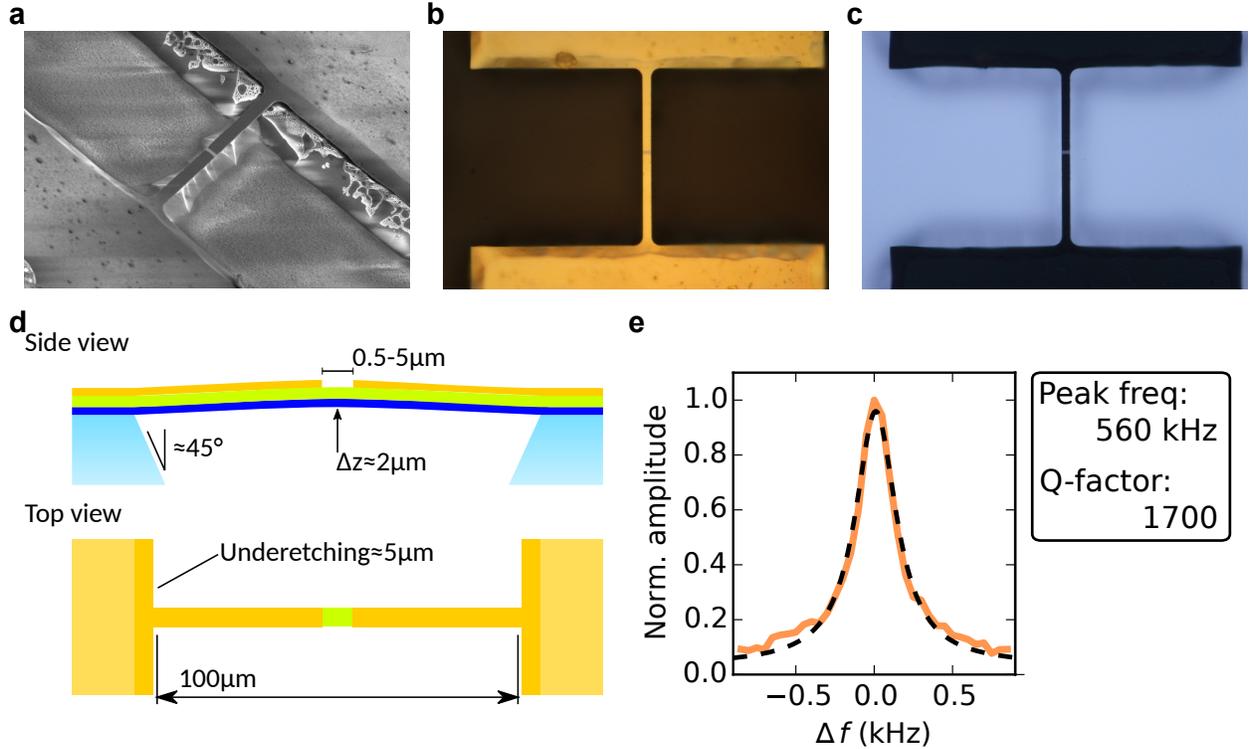

Figure S2. Final device: (a) SEM image, (b) optical image in reflected light, (c) optical image in transmitted light, (d) schematic of the device profile and (e) resonance peak (orange line) and Lorentzian fit (black dashed line) of one of the lowest mechanical modes of the structure at 30 °C.

Figure S2 shows images of a typical device acquired with different techniques, together with the peak shape of one of the mechanical modes of the resonator and a schematic showing the microbridge profile. Due to the built-in strain coming from the lattice mismatch between $MgO/TiO_2/VO_2$, our microbridges are slightly buckled. We estimated the height difference between the edge of the structure and its centre using a calibrated optical microscope, resulting in a height difference of about 2 $\mu$m over a 100 $\mu$m-long microbridge. Due to the anisotropic etching rate of MgO, the edges in the under-etched regions show preferential 45° faceting, corresponding to the (011) and (101) planes. In panel e we show the peak-shape of one of the lowest mechanical modes of the micro-resonator measured at low temperature (30 °C). By fitting with a Lorentzian function the resulting Q-factor is about 1700, fairly high if considering the presence of different materials and the non-ideal geometry.

## III. Shift of the mechanical modes under current and temperature bias

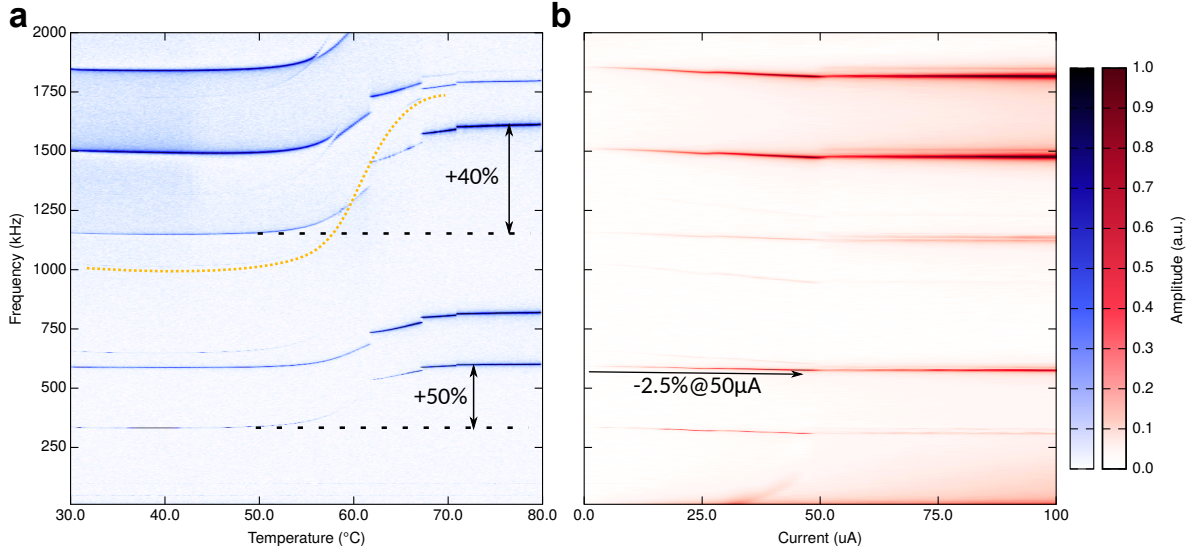

Figure S3. Colormaps of the mechanical spectra of the microbridge of Fig.1a as a function of temperature (**a**) and DC current bias (**b**).

Figure S3a shows the shift of the resonance peaks to higher values with increasing temperature. Their relative variation is about $40\% - 50\%$, depending on the specific mode, in agreement with previous reports (see Ref. 26). This shift is mainly due to the stress-stiffening effect, which increases the total rigidity of the structure upon accumulation of stress followed by the lattice transition. In our device this effect is suppressed when biasing by an electrical current. In Figure S3b we show the mechanical spectra of the microbridge for bias currents up to $100\ \mu A$, measured at $30\ °C$. The values of the resonance peaks slightly decrease up to $50\ \mu A$ and is constant at higher values. This flat response, even at current values where the electrical resistance has already dropped due to the metallization of $VO_2$, indicates that the Ti/Au electrodes are successfully confining the phase transition between the gap.

The initial decrease in eigenfrequency of Figure S3b cannot be due to the accumulated stress induced by the phase transition of $VO_2$, since it would cause an increase of the eigenfrequencies, as discussed in detail in Figure S3a. The frequency shift downward is attributed to Joule heating, which causes a reduction of the Young's modulus. The flat response above $50\ \mu A$ can be easily explained by considering Figure 1e, where above $40\ \mu A$

the electrical resistance stops increasing. This causes a stabilization of the dissipated Joule power, limiting the temperature increase and the associated mechanical frequency shift.

We note that the mechanical spectra also show peaks with small intensity. These peaks show a wider relative variation with temperature and one of them is indicated by an orange dotted line in Fig. S3a. The presence of these modes becomes clearer when they cross other resonances, determining a mode splitting. The lower amplitude of these resonances and their different temperature dependence, if compared to the most visible ones, could indicate that these peaks might be due to torsional modes.

## IV. Strain induced by the VO$_2$ phase transition

The TiO$_{2/}$VO$_2$ heterostructure is grown by pulsed laser deposition on top of a MgO(001) single-crystal substrate. The TiO$_2$ layer grows with the (110) planes perpendicular to the c-axis of the substrate, while its [001] direction can be oriented both along [110] and [1$\bar{1}$0] directions of the MgO(001), creating a pattern of orthogonal nanodomains [1]. VO$_2$ grows epitaxially on top of the TiO$_2$ and shows only ($ll$0) peaks [2, 3], meaning that also VO$_2$ grows with two in-plane orientations of strained nanodomains. It is possible to calculate the strain applied by the phase transition of VO$_2$ by considering these growth directions and the differences in the unit cell size. The table below shows the dimensions of the unit cell of VO$_2$ bulk in the different phases.

| Phase | a[Å] | b[Å] | c[Å] | $\beta[deg]$ | Ref. |
|-------|------|------|------|--------------|------|
| R | 4.5546 | 4.5546 | 2.8514 | - | [4] |
| M$_1$ | 5.7517 | 4.5378 | 5.3825 | 122.65 | [5] |
| M$_2$ | 9.060 | 5.800 | 4.5217 | 91.85 | [6] |

Because of the different definition of a,b,c in the monoclinic and rutile phases it is not possible make a direct comparison. The following table shows the approximated equivalent directions between the different phases.

| $R \rightleftharpoons M_1$ | | |
|------|------|------|
| $a_{M_1}$ | $b_{M_1}$ | $c_{M_1}$ |
| $2c_R$ | $b_R$ | $\sqrt{a_R^2 + c_R^2}$ |

Considering the two possible in-plane orientations of the [110]-oriented VO$_2$ domains, the total in-plane strain due to the phase change is isotropic and its magnitude is given by the average over the two directions:

$$\epsilon = \frac{(\sqrt{a_R^2 + c_R^2} + b_R) - (c_{M_1} + b_{M_1})}{(c_{M_1} + b_{M_1})} \approx 0.08\%$$

## V. Finite elements simulations of the eigenmodes of the structure

In order to understand how the buckling of the structure and the strain induced by the phase transition of $VO_2$ can affect the mechanical properties of the microbridge, we performed finite elements simulations of the mechanical modes of the structure. The analysed geometry is a 2D longitudinal section of the microbridge profile. This provides a simple model of the interplay between strain and mechanical modes of this structure, although it prevents to simulate the transversal modes and eventual deformations in the transversal direction. The simulated bridge is 100 $\mu$m-long with an effective width of 5 $\mu$m and a 2 $\mu$m gap between the metallic electrodes. The materials properties of the $TiO_2/VO_2/Au$ heterostructure are presented in the following table:

|        | Density [$kg/m^3$] | Young's Modulus [$GPa$] | Poisson's ratio |
|--------|--------------------|--------------------------|-----------------|
| Au     | 19300              | 70                       | 0.44            |
| $TiO_2$ | 4230              | 282                      | 0.28            |
| $VO_2$ | 4670               | 140                      | 0.3             |

We performed a parametric simulation applying a compressive (positive) strain to the $VO_2$ layer from 0% to 0.22% in the two in-plane directions. These values represent what observed experimentally: for a strain of 0.14% the resulting buckling in the centre of the microbridge is about 2 $\mu$m, similarly to what observed in Figure S2, then a further 0.08% of strain represent the effect of the $VO_2$ phase transition, as discussed in the Supplementary section IV. To break the symmetry of the system and determine a positive buckling, a small homogeneous force of 100 nN is applied to the bottom edge of the structure, which results in an initial deflection of about 100 nm. The results of this analysis are presented in Figure S4. Figure S4a shows the frequency of the first eight longitudinal mechanical eigenmodes as a function of the strain. We note that, because of the small initial force and the presence of the gap, their distribution does not match with what expected for an ideal beam. The black dashed line indicates the 0.14% strain value, while the red one is at 0.22% (+0.08%). The buckled profile of the microbridge for a 0.14% strain is shown in Figure S4b, where the inset is a magnification of the bridge centre. The mode shape calculated at this buckling condition are presented in Figure S4c, where the order is given by their eigenfrequency value in this strain condition.

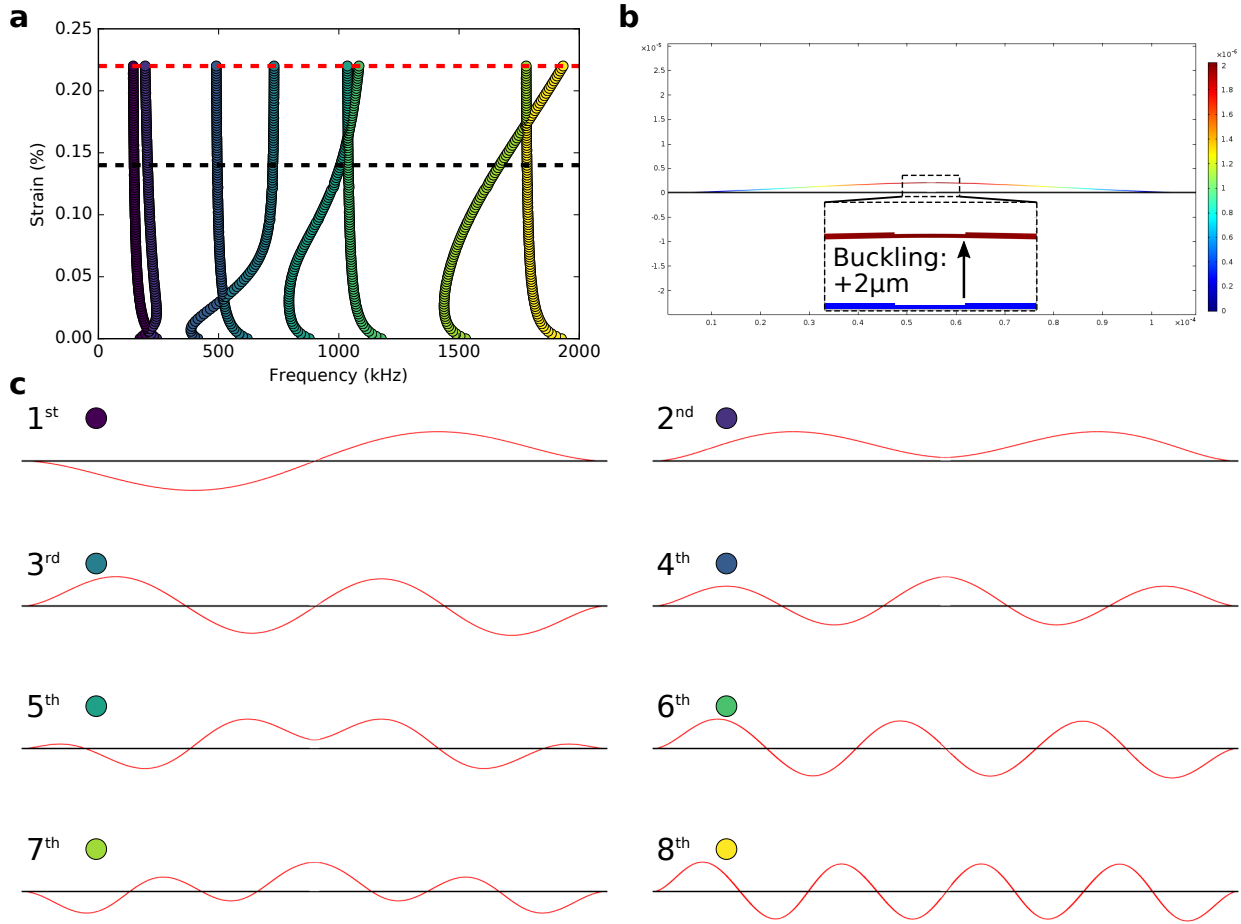

Figure S 4. Finite elements simulation of the longitudinal mechanical modes of a 100 $\mu$m TiO$_2$/VO$_2$/Au buckled microbridge under compressive strain. (a) Frequency of eigenmodes as a function of the strain. The black dashed line is the value at which a buckling of 2 $\mu$m is reached (0.14%), the red dashed line correspond to a further increase of 0.08%. (b) 2D profile of the microbridge with a strain of 0.14%,resulting in a deflection of 2 $\mu$m, as observed on the real device at room temperature. The inset shows a magnification of the microbridge centre with the initial (blue) and buckled (red) profile. (c) Mode shape of the first eight eigenmodes calculated at 0.14% strain (black dashed line of (a)), the order is given by their eigenfrequency value.

From these simulations it is evident how the spectral distribution of the normal modes is strongly dependent on the applied strain. This results in two effects that partially balance each other. From one side an increased strain results in an accumulated compressive stress that makes the structure less rigid, reducing the frequencies of the mechanical modes. On the other side, the strain is progressively relaxed by the buckling of the bridge that increases

10

the rigidity of the structure [7]. This is manifested by the different dependence from the strain of the modes, where some of them are strongly strain-dependent with non-monotonic behaviour, while others shows a smaller response. Although the results of Figure S4a provide an explanation of several features of the mechanical response of our structure, such as the non-ideal mode distribution and the mode-crossing observed in Figure S3, it still can not fully describe the high frequency shift observed during a temperature sweep. Our hypothesis is that this characteristic is determined by the geometric stiffening effect, i.e. a deformation of the microbridge shape that results in a large increase of its rigidity. Its presence is expected considering that the strain at the phase transition is isotropic, meaning that it can also determine a curvature of the microbridge along its transversal direction. This type of geometric distortion can result in a large increase of the rigidity of the structure [8], meaning that it probably represents a relevant contribution to the observed frequency shift.

## VI. Optical lever setup and sample holder

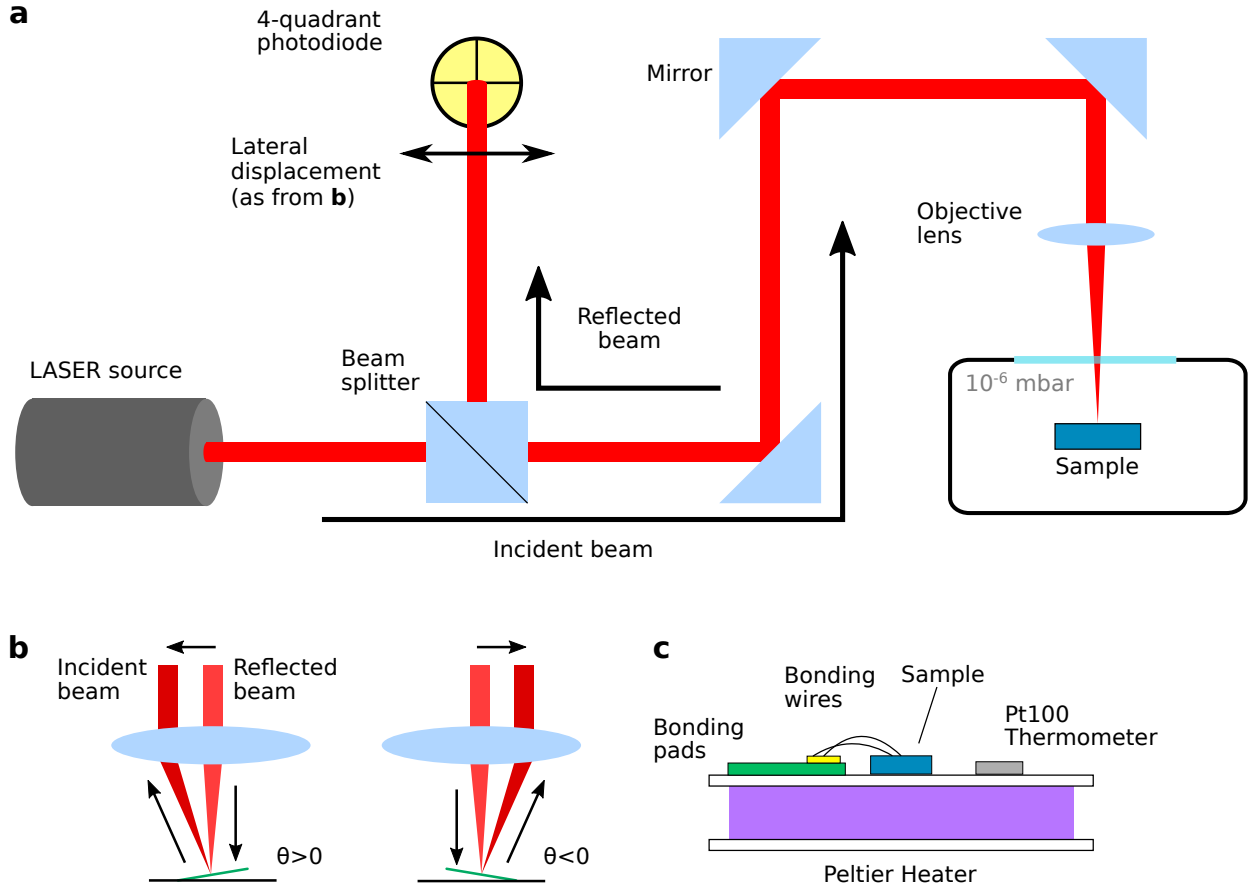

Figure S5. Measurement setup: (a) Diagram of the optical lever geometry used to measure the microbridge motion. (b) Schematic of the working mechanism of the optical lever. (c) Schematic side-view of the sample holder.

Figure S4a shows a diagram of the optical lever setup used to measure the motion of the microbridges. A laser beam is focused on the top surface of the structure using a series of mirrors and an objective lens. The motion of the microstructure produces a tilt of its surface changing the direction of the reflected beam (figure S4b). The reflected beam follows the same optical path of the incident one up to the initial beam splitter that sends it to a four-quadrant photodiode which converts the shift of the laser beam to an electrical signal. The sample is mounted on top of a Peltier element with a local Pt100 thermometer (figure S4c) placed in a vacuum chamber with base pressure of about $10^{-4}$ mbar. The single devices are connected with an ultrasonic wire bonder to apply the electrical bias.

## VII. Contributions to the electro-thermal actuation

### A. Monoclinic-rutile transition of the $VO_2$ within the gap

To evaluate the magnitude of the periodic expansion and contraction of the $VO_2$ between the Au electrodes due to the electrical oscillations, we performed a finite elements simulation of the static displacement in the two conditions. The simulated geometry was a 100 $\mu$m-long microbridge with a gap between the metal electrodes of 2 $\mu$m. The different materials were modelled as discussed in Supplementary section IV. Initially we applied a built-in strain to the whole structure to reach a vertical deflection of about 2 $\mu$m, as measured by the optical microscope in Figure 2b, then we calculated the vertical displacement of the microbridge while increasing the built-in strain to the $VO_2$ between the electrodes. The result is presented in Figure S6, where for a built-in strain of 0.08 % we observe a vertical shift of about 32 nm. The range of calculated strain was chosen by taking into account the discussion of Supplementary section IV. The final displacement is rather high, since it corresponds to about the 20 % of the total thickness of the heterostructure. Although the precision of this modelling may be limited by the assumptions of homogeneous media and

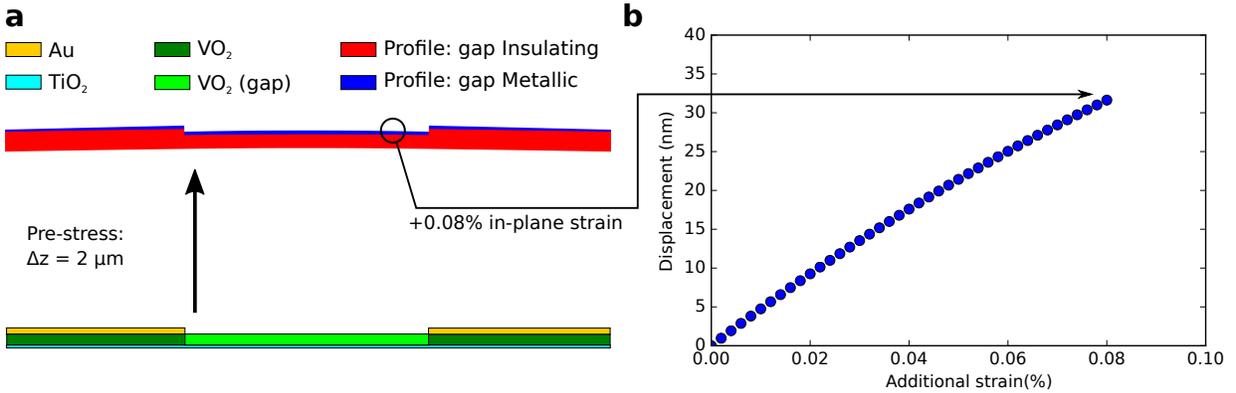

Figure S 6. Static actuation from the electrical-induced phase transition by Finite elements Simulations. (a) Profiles of the buckled microbridge: the initial heterostructure is pre-stressed to reach a buckling condition of about 2 $\mu$m Static displacement of a 100 $\mu$m microbridge when the $VO_2$ between the electrodes (2 $\mu$m gap) is in the insulating monoclinic phase (red) or in the metallic rutile phase (blue). (b) vertical displacement as a function of the strain of the $VO_2$ between the electrodes.

temperature-independent coefficients, it indicates that the strain applied by the $VO_2$ during the electrical oscillations is a relevant contribution to the observed mechanical actuation

## B. Periodic thermal expansion from Joule heating

A rough evaluation of this contribution can be provided by comparing the amplitude of the oscillating current during the electrical oscillations and what we used to excite the microstructure to characterize its mechanical properties. From Figure 2b, the amplitude of the oscillations of the electrical current is more than 15 $\mu$A. This is weakly affected by value of $V_0$ or $R_0$, at least in the range of values used in this experiment. This is more than three times higher than the AC current used for thermomechanical excitation of the structure during the measurement of the the mechanical spectra of the microresonator in Figure 1c and Figure S3, where 5 $\mu$A were applied. For a proper comparison we need to estimate the dissipated Joule power, which is directly related to the temperature increase and thus the thermal expansion. During the mechanical characterization the gap is insulating and the gold resistance is negligible, according to the resistance value measured in Figure 1d, the sample resistance at room temperature is about 50 k$\Omega$. The resulting average Joule power is:

$$P_{\text{VNA}} = \frac{1}{2} \times 5 \cdot 10^4 \, \Omega \times (5 \cdot 10^{-6} \, A)^2 \approx 0.62 \, \mu W$$

For the electro-thermal oscillations we consider the average of the Joule power dissipated in the $VO_2$ between the insulating and metallic phase. From Figure 2b the magnitude of the electrical current oscillate between 60 $\mu$A in the insulating state and 80 $\mu$A in the metallic state. At the same time the voltage drop across the microbridge varies between 0.6 V in the insulating state and almost 0 V in the metallic state. The average oscillating Joule power dissipated in the microbridge can be approximated as:

$$P_{\text{EO}} = \frac{1}{2} \times 60 \, \mu A \times 0.6 V \approx 18 \, \mu W$$

To compare the thermomechanical excitation applied by the VNA during the device characterization with the power dissipated during the electrothermal oscillations, it must be considered that in the first case the current waveshape is sinusoidal while in the latter double-exponential. This means that the spectral power density of the two excitations are different: $J_{\text{VNA}}$ is always applied at a single (sweeping) frequency, while $J_{\text{EO}}$ is spread across

multiple harmonic components and just the one that is in the matching condition with a flexural mode is able to provide mechanical excitation.

## C. Capacitive force from the oscillating potential of the microbridge

The periodic force coming from the electrostatic coupling is due to the oscillating electrical potential of the structure with respect to ground. The capacitive force depends from the square of the voltage ($F_C \propto V^2$) this operation does not preserves the frequency and for a signal with zero average, results in a frequency doubling:

$$F_C \propto (V_0 \cos(\omega_0 t))^2 = \frac{V_0^2}{2}(1 + \cos(2\omega_0 t)$$

However, if the voltage oscillation does not average to zero, a component at $\omega_0$ is preserved:

$$F_C \propto (V_0(a_0 + \cos(\omega_0 t))^2 = \frac{V_0^2}{2}((2a_0 + 1) + 4\cos(\omega_0 t) + \cos(2\omega_0 t))$$

which is precisely the case of the electrical oscillations in $VO_2$, where, from Fig. 3c, the electrical voltage oscillates from 0 V to 0.6 V. To evaluate the resulting force applied on the microstructure, we can consider a simple capacitive model where one plate is the microbridge and the other the (grounded) bottom face of the 0.5 mm-thick MgO substrate ($\epsilon_r = 10$). We did a simple finite elements simulation, considering a 100 $\mu$m bridge at 0.6 V with a 10 $\mu$m air gap below. The resulting total force is about 0.2 nN, giving a bridge displacement of 1 Å, calculated using the same model of Supplementary section V and VII.

## Summary

In this section we estimated the magnitude of the different contributions to the mechanical actuation during the electro-thermal oscillations. The lattice-induced strain and the Joule-assisted heating seems both able to provide a relevant mechanical actuation alone, while the electrostatic coupling appears to be weaker if compared to them. Future experiments, with dedicated device geometry like side gates or different aspect-ratio between the $VO_2$ gap and the bridge length, may allow to experimentally investigate, and eventually control, the relative magnitude of these contributions.



\end{document}